\documentclass{pasa}%

\usepackage{graphicx,epsfig,subfigure}
\usepackage{amssymb,amsmath}
\usepackage{epstopdf}

\title[Comparison of observing modes for statistical estimation of the EoR signal]{Comparison of observing modes for statistical estimation of the 21~cm signal from the Epoch of Reionisation}
\author[C. M. Trott]{Cathryn M. Trott$^{1,2}$\thanks{cathryn.trott@curtin.edu.au}\\
\affil{$^1$International Centre for Radio Astronomy Research, Curtin University, Bentley Australia}%
\affil{$^2$ARC Centre of Excellence for All-Sky Astrophysics (CAASTRO)}}%
\jid{PASA}
\doi{10.1017/pas.\the\year.xxx}
\jyear{\the\year}

\usepackage[authoryear]{natbib}
\bibpunct{(}{)}{;}{a}{}{,}
\begin{document}%
\begin{abstract}
Noise considerations for experiments that aim to statistically estimate the 21~cm signal from high redshift neutral hydrogen during the Epoch of Reionisation (EoR) using interferometric data are typically computed assuming a tracked observation, where the telescope pointing centre and instrument phase centre are the same over the observation. Current low frequency interferometers use aperture arrays of fixed dipoles, which are steered electronically on the sky, and have different properties to mechanically-steered single apertures, such as reduced sensitivity away from zenith, and discrete pointing positions on the sky. These properties encourage the use of two additional observing modes: (1) zenith drift, where the pointing centre remains fixed at the zenith, and the phase centre tracks the sky, and (2) drift$+$shift, a hybrid mode where the telescope uses discrete pointing centres, and the sky drifts during each fixed pointing. These three observing modes view the sky differently, and therefore yield different uncertainties in the power spectrum according to the balance of radiometric noise and cosmic variance. The coherence of measurements made by the instrument in these modes dictates the optimal reduction in thermal noise by combination of coherent modes, and the reduction in cosmic variance by combination of incoherent modes (views of different patches of the sky). Along with calibration and instrument stability considerations, the balance between these noise components provides one measure for the utility of these three modes for measuring a statistical signature of the EoR signal. We provide a general framework for estimating the uncertainty in the power spectrum for a given observing mode, telescope beam shape, and interferometer antenna distribution. We then apply this framework to the Murchison Widefield Array (MWA) using an analysis of the two-dimensional (2D) and one-dimensional (1D) power spectra for 900 hours of observing. We demonstrate that zenith drift scans can yield marginally lower uncertainty in the signal power compared with tracked scans for the MWA EoR experiment, and that moderately higher signal-to-noise ratio (S/N) estimates of the amplitude (3\%) and slope (1\%) of the 1D power spectrum are accessible, translating directly into a reduction in the required observing time to reach the same estimation precision. We find that the additional sensitivity of pointing at zenith, and the reduction in cosmic variance available with a zenith drift scan, makes this an attractive observing mode for current and future arrays.
\end{abstract}
\maketitle%
\section{INTRODUCTION }
\label{sec:intro}

Measuring signals from the early Universe is a major goal of many current and future low-frequency radio telescopes \citep[MWA, PAPER, LOFAR, SKA, LEDA, HERA\footnote{http://www.reionization.org}, ][]{tingay13_mwasystem,parsons10,stappers11,greenhill12_leda,vanhaarlem13}. The Epoch of Reionisation is the period in the early Universe marking the end of the Dark Ages, when the first sources of radiation ionised the neutral intergalactic medium. The evolution of structure before and during this period can be probed by observing the neutral hydrogen emission line at rest frequency of 1420~MHz, because it traces the distribution and state of hydrogen as a function of redshift. At the redshifts where the EoR signal is expected to be detected ($z\sim{6-10}$), this transition is observable with low-frequency instruments ($\nu\sim{130-200}$~MHz).

The signal is predicted to be weak compared with other radio sources at low frequencies; temperature fluctuations against the Cosmic Microwave Background are expected to be $\sim$10s~mK, while emission from foreground continuum sources (e.g., Galactic synchrotron, diffuse free-free emission, supernova remnants, radio galaxies) can exceed 100s~K \citep{pritchard08,jelic10,mesinger11}. Careful discrimination of spectral line signal from the early Universe from contaminating sources relies on the distinct spatial and spectral structure for these sources; radio galaxies and other point source-like foregrounds contribute power on small angular scales, while the EoR signal is expected to peak on larger scales (a few degrees). Diffuse emission from the Galaxy, in particular in the Galactic plane, are spatially similar to this signal \citep{jelic10}. Conversely, the spectral structure of foreground continuum sources versus spatially-variable spectral line emission from the EoR provides a strong discriminator for the signals. Coupled with the weakness of the expected EoR signal, and the need to combine information from across the sky (stack the signal to increase detectability), statistical estimates are a useful tool for EoR detection and estimation. In particular, the two-dimensional power spectrum (signal variance as a function of angular and line-of-sight mode) is a useful metric for discriminating EoR signal from foregrounds on the basis of spatial and spectral differences, and for increasing detectability by combining signal statistically \citep{morales04,mcquinn06}. Measuring the statistical EoR signal relies on (1) a stable and calibratable instrument, and (2) sufficient data to reduce thermal noise, measure and remove foregrounds, and reduce cosmic variance \citep{mellema13}.

Statistical measurement of the EoR signal has many challenges, including reduction of noise to yield a detection. The two primary sources of noise for a stable and well-calibrated instrument (i.e., one where radiometric noise exceeds calibration uncertainties and calibration errors are minimal) are radiometric (stochastic measurement noise because we are sampling the signal) and cosmic variance (uncertainty in measurement of a stochastic global signal because it is estimated by finite data). The most effective methods for reducing these noise contributions are not the same.

The expected EoR signal power is low at very large angular scales, with the peak power occurring on scales of $\sim$~2 degrees \citep[see ][and references therein]{mellema13}. Large scales are sampled by the shortest baselines of an interferometer, and most arrays have few. Therefore, at low $k$, we expect thermal noise to dominate. At larger $k$, where the EoR signal is expected to be strongest, and baselines are numerous (by design for an EoR experiment), cosmic variance becomes relatively more important. The balance between these coherent and incoherent modes is the subject of this note.

Noise analysis has been performed before to predict the performance of telescopes \citep{morales04,mcquinn06,parsons12,beardsley13,mellema13}, but these have typically used an approximate form for the noise estimate. In this work we build on the existing literature by: (1) deriving a framework that correctly accounts for the telescope's beam pattern \citep[and therefore the level of coherence/incoherence in a measurement set. See][who also considered this aspect]{thyagarajan13}; (2) including covariances between angular modes in the calculation, and (3) applying it to the specific case of comparing different observing modes. In their work, \citet{thyagarajan13} considered both tracked scans and zenith scans with independent pointing centres. \citet{parsons12} also considered drift scans for exploring observing options for PAPER. We consider a scan with a constant evolution of the pointing centre, as the Earth rotates, requiring a robust framework to capture the time-dependent coherence of two baselines relative to each other.

Traditional tracked observing modes, where the pointing and phase centres align and track a single sky position, are undesirable for aperture arrays with analogue beamformers, where such modes produce a time-variable sky beam because the individual antenna elements have discrete timing offset settings (`delays'). Instead, two modes can be employed, which have different calibration and EoR measurement properties: (1) a ``zenith drift", where the pointing centre remains fixed at the zenith, and the phase centre tracks a single position on the sky, and (2) ``drift$+$shift", a hybrid mode where the beam has fixed, discrete pointing centres, with the sky drifting during each fixed pointing. While providing a stable, constant beam, the former sees a time-variable sky, thereby reducing the \textit{coherence} of measurements across a long observation. I.e., fewer individual visibility measurements sample the \textit{same} sky, at the \textit{same} baseline vector. For EoR statistical estimation, sample coherence is used to combine visibilities directly. Coherent addition is desirable because thermal noise is reduced the most effectively. On the other hand, cosmic variance does not benefit from redundant samples of the same sky; coherent sampling maintains cosmic variance. Measurements that sample the same modes on the sky, but at different orientations or pointing in different directions, can be added incoherently (powers added). This has less benefit for reducing thermal noise, but \textit{does} reduce cosmic variance (independent samples of the sky).



The Murchison Widefield Array (MWA) is a low-frequency (80--300~MHz) radio telescope composed of 128 tiles operating in the Western Australian desert \citep{tingay13_mwasystem}. It has a concentrated core of 50 tiles within $\sim$300~m, and the remainder of the tiles spread over an area of diameter $\sim$3~km. Each tile consists of 16 phased dual-polarisation dipoles in a 4-by-4 grid. One of the MWA's primary science goals is statistical detection of the EoR in the redshift range $z\sim 6-10$ \citep{bowman13_mwascience}. In the current semester it is observing two EoR fields away from the Galactic plane, and in relatively cold patches of the sky, over two frequency ranges (139--169~MHz, 167--197~MHz) with 40~kHz spectral resolution. The initial experiment will be based on two fields with a combined observing time of $\sim$1000 hours. \citet{beardsley13} estimate that a 14$\sigma$ detection is possible in the spherically-averaged power spectrum for this experiment.

In section \ref{ml_estimate} we describe the power spectrum, and its derivation from an interferometric dataset. We then connect the measured data to the information it provides about the temperature fluctuations in a given sky mode, and use this model to propagate the noise terms to the final power spectrum. In section \ref{beam_section} we describe the differences in the noise estimates for each observing mode, based on the way in which the telescope's beam pattern samples the sky. We then apply the model to a simulation of the EoR experiment for MWA in section \ref{mwa_experiment}. Throughout we use comoving co-ordinates in a $\Lambda$CDM cosmology, with $\Omega_m=0.27$, $\Omega_\Lambda=0.73$, $\Omega_k$=0.0, and $H_0 = 100h$ \citep{bennett12}. Matrices are denoted in bold font, and vectors are denoted by an overline.
 
\section{Methods}
\label{ml_estimate}
The power at a given scale (a ``mode") is given by:
\begin{eqnarray}
P(\overline{u})\delta_{uu^\prime} = \left\langle S^\dagger(\overline{u}) S(\overline{u}^\prime) \right\rangle = \left\langle \Delta\tilde{T}^\dagger(\overline{u}) \Delta\tilde{T}(\overline{u}^\prime) \right\rangle,
\end{eqnarray}
where ${S}({\overline u})$ is the complex-valued Fourier transform of the sky signal and traces temperature fluctuations, $\Delta{\tilde{T}}(\overline{u})$, the ensemble average is taken over different samples from the sky, and the delta function has units of volume and connects the integrated sky power (K$^2$.sr.Hz) with the integrated sky signal (K.sr.Hz).

The aim of this paper is to quantify the information our data contain about the power in mode $u$, $P(u)$. Typically, one would write a description of the forward model connecting the sky with the measurements, and compare the noise properties of the measurement set. However, because the number of measurments (visibilities) is very large compared with the number of sky modes in which we are interested (millions per day compared with hundreds for the whole observation), it is prohibitive to describe the full covariance in the measurement basis. Instead, we explore the inverse problem, where we describe the information available in a particular sky mode due to the underlying measurement set. Working in this basis provides a simpler framework with which to tackle the problem.

\subsection{Coherent average visibility in cell $\overline{u}$}
The true sky signal is shaped by the antenna primary beam, and the measured visibilities sample a range of modes within this beam-attenuated sky. Using the convolution theorem and considering the measured signal to be a two-dimensional Fourier transform, the measured visibility is the convolution of the sky signal, $S(\overline{u})$, with the Fourier representation of the primary beam and the line-of-sight ($w$) term, $B(\overline{u})$. Here we approximate the continuous transform by a discrete transform:
\begin{eqnarray}
V(\overline{u}) &=& {B}(\overline{u}) \ast S(\overline{u}) + N\\
&=& \displaystyle\sum_{{\overline u}^\prime} B(u-u^\prime)S(\overline{u}^\prime) + N.
\end{eqnarray}
$N$ is the radiometric noise, which is white Gaussian distributed in the measurement basis. $B$ encodes the Fourier-space primary beam shape and the discrete telescope sampling of modes (the $uv$ coverage), and the smearing of sky signal by non-zero $w$ terms,
\begin{equation}
B(\overline{u}) \equiv \displaystyle\int \frac{B(l,m)}{n}\exp{-2\pi{i}(ul+vm+w(n-1))}dldm,
\end{equation}
where $n=\sqrt{1-l^2-m^2}$.
We are interested in determining the underlying sky distribution, $S$, at a given angular scale, $u$, given the measured visibilities, $V(\overline{u})$. A vector of ($N\times{1}$) measured visibilities, $\overline{V}$, is generated from an underlying ($M\times{1}$) sky distribution, $\overline{S}$, via a ($N\times{M}$) matrix transform, $\boldsymbol{B}$, such that:
\begin{equation}
\overline{V} = \boldsymbol{B}\overline{S},
\end{equation}
where $M{\ll}N$, in general. The beam matrix, $\boldsymbol{B}$, is complex-valued, in general.

To form the coherent average of measured visibilities into modes $\overline{u}$, we solve for $\overline{S}(\overline{u})$,
\begin{equation}
\overline{S}(\overline{u}) = (\boldsymbol{B}^\dagger\boldsymbol{B})^{-1}{\boldsymbol{B}}^\dagger \overline{V},
\label{beam_inverse}
\end{equation}
where the square matrix inversion involves inverting an $(M\times{M})$ Hermitian complex-valued matrix, which will be (almost) diagonal for independent modes, $\overline{u}$. Note that this procedure does not recover full sky information; the sampling function of the interferometer does not provide complete $uv$-coverage, and the problem is under-constrained. Instead, this procedure distributes the measured visibility data across contributing sky modes, to represent the smearing of Fourier space by the primary beam (grids and weights). It also represents a continuous transform with a discrete approximation, which destroys some information.


We do not measure $S(\overline{u})$, but $V(\overline{u})$, visibilities that include beam contributions and noise. The visibility is given by:
\begin{eqnarray}
V(\overline{u}) = V^{21}(\overline{u}) + V^{F}(\overline{u}) + N,
\end{eqnarray}
where the terms are the 21~cm signal, foreground signal and noise, respectively, and
\begin{equation}
N \sim \mathcal{CN}(0,\sigma^2),
\end{equation}
describes the radiometric noise with variance, $\sigma^2$.
If we consider the covariance matrix of the data, we find;
\begin{eqnarray}
\boldsymbol{C}_{\rm d} &=& \boldsymbol{C}^{21} + \boldsymbol{C}^{F} + \boldsymbol{C}^{N}\\
&\equiv& \left\langle V^{21}(\overline{u}) V^{21\dagger}(\overline{u}^\prime) \right\rangle + \left\langle V^{F}(\overline{u}) V^{F\dagger}(\overline{u}^\prime) \right\rangle  + \sigma^2\boldsymbol{I},
\end{eqnarray}
where the first two terms correspond to the cosmic variance of the 21~cm signal and foregrounds, and the final term is the thermal noise contribution. In the analysis that follows we will retain the contribution from foreground signal as a separate term, and focus on the thermal noise and cosmological signal terms. As a signal, foreground contributions exhibit the same behaviour as cosmological signal (cosmic variance noise considerations). We will assume herein that the foregrounds have been removed to a certain level, but their residuals (which may be significant compared with the signal of interest) remain in the data. In the final analysis of the estimation performance of each method, we will consider uncontaminated regions of parameter space. For this work, where we focus on small $k$ modes (short baselines), this will correspond to omitting the $k_{\parallel}$=0 mode in the power spectrum. We revisit the impact of foregrounds in sections \ref{foregrounds} and \ref{discussion}. An excellent recent discussion of the impact of foregrounds can be found in \citet{thyagarajan13}.

\subsection{Coherent combination of information: sky covariance}


For each visibility, we compute the contribution to the sky at mode $u$:
\begin{equation}
S(\overline{u}) = \left( \boldsymbol{B}^\dagger \boldsymbol{B} \right)^{-1} \boldsymbol{B}^\dagger V(\overline{u}^\prime).
\end{equation}
We form a grid of sky modes so as to properly sample the beam shape for imaging a large sky angle (necessary for the drift scan). This choice ensures coherence between adjacent modes for all observing strategies (the same visibilities contribute significantly to both modes). To determine the effective number of visibilities contributing coherently to a mode, and the covariance between modes, we compute the outer product of the sky signal in each mode (sky covariance matrix), given the information about that mode provided by the instrument:
\begin{eqnarray}
\label{c_s}
\boldsymbol{C}_s &\equiv& \left\langle S(\overline{u_i}) S^{\dagger}(\overline{u_j}) \right\rangle  \\\nonumber &=& \left( \boldsymbol{B}^\dagger \boldsymbol{B} \right)^{-1} \boldsymbol{B}^\dagger \left\langle V(\overline{u_i}) V^{\dagger}(\overline{u_j}) \right\rangle \boldsymbol{B} \left( \boldsymbol{B}^{\dagger}\boldsymbol{B} \right)^{-1} \\\nonumber
&=& \left( \boldsymbol{B}^\dagger \boldsymbol{B} \right)^{-1} \boldsymbol{B}^\dagger \sigma^2 I \boldsymbol{B} \left( \boldsymbol{B}^{\dagger}\boldsymbol{B} \right)^{-1} + \boldsymbol{P}^\prime + \boldsymbol{C}_F \\\nonumber
&=& \sigma^2 \left( \boldsymbol{B}^{\dagger}\boldsymbol{B} \right)^{-1}_{ij} +  \boldsymbol{P}^\prime  + \boldsymbol{C}^{F^\prime},
\end{eqnarray}
where
\begin{eqnarray}
\boldsymbol{P}^\prime &\equiv& \left( \boldsymbol{B}^\dagger \boldsymbol{B} \right)^{-1} \boldsymbol{B}^\dagger \\\nonumber
&& \left\langle {\iint B(\overline{u}-\overline{u}^\prime) B^\ast(\overline{v}-\overline{v}^\prime) S_{21}(\overline{u}^\prime) S_{21}^\ast(\overline{v}^\prime) d\overline{u}^\prime d\overline{v}^\prime} \right\rangle \\\nonumber
&& \boldsymbol{B} \left( \boldsymbol{B}^{\dagger}\boldsymbol{B} \right)^{-1} \\\nonumber
&=& \left( \boldsymbol{B}^\dagger \boldsymbol{B} \right)^{-1} \boldsymbol{B}^\dagger \\\nonumber
&& {\iint B_1(\overline{u}-\overline{u}^\prime) B_2^\ast(\overline{v}-\overline{v}^\prime) P(\overline{u}^\prime) \delta(\overline{u}^\prime-\overline{v}^\prime) d\overline{u}^\prime d\overline{v}^\prime} \\\nonumber
&& \boldsymbol{B} \left( \boldsymbol{B}^{\dagger}\boldsymbol{B} \right)^{-1} \\\nonumber
&=& \rm{diag}\left( \boldsymbol{B}^\dagger \boldsymbol{B} \right)^{-1} \displaystyle\int_{u^\prime} B_1(\overline{u}-\overline{u}^\prime) B_2^\ast(\overline{v}-\overline{u}^\prime) P(\overline{u}^\prime) d\overline{u}^\prime \\\nonumber
&\approx& \rm{diag}\left( \boldsymbol{B}^\dagger \boldsymbol{B} \right)^{-1} {\boldsymbol{P}(\overline{u})} \displaystyle\int_{u^\prime}  B_1(\overline{u}-\overline{u}^\prime) B_2^\ast(\overline{v}-\overline{u}^\prime) d\overline{u}^\prime \\\nonumber
&\equiv& {P(\overline{u}_i)}{\Omega_u}.
\end{eqnarray}
Here, $\Omega_u$ is an effective volume in Fourier space, which is equal to the beam area for a tracked scan ($B_1=B_2$ for the whole observation). Consider what has occurred here: the instrument convolves the primary beam shape with the intrinsic sky signal (in the continuum limit) and samples this convolution at discrete points in the $uv$-plane. If the beam pointing changes during the observation ($B_1 \neq B_2$), the two samples of the sky are more independent and the measurement-space covariance matrix is reduced in amplitude, thereby reducing the contribution to the total noise budget of cosmic variance. The normalisation by the diagonal components of the $\left( \boldsymbol{B}^\dagger \boldsymbol{B} \right)^{-1}$ matrix removes (normalises) the improvement one would otherwise compute by sampling the sky coherently (i.e., sampling the same sky signal does not improve cosmic variance).

\begin{figure}
\begin{center}
\includegraphics[scale=0.35]{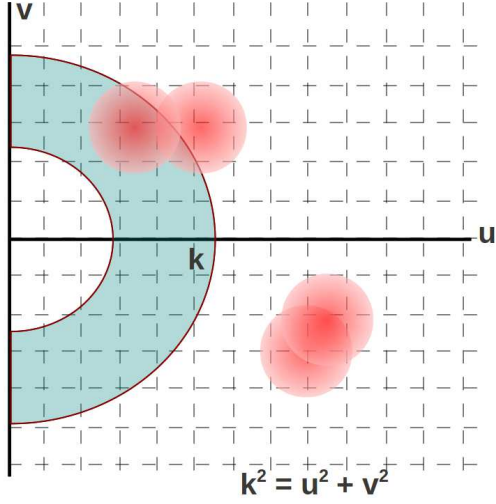}
\caption{Schematic representation of the $uv$-plane (Fourier plane), and the footprint of the primary beam (station beam) from individual visibility measurements (red circles). Also shown in the blue annular region is an example area over which the cosmological signal is not expected to evolve substantially, and the powers are averaged incoherently. The finite size of the beam in Fourier space mixes information from different sky modes into an individual visibility measurement.}\label{beam_footprint}
\end{center}
\end{figure}

For an ideal instrument, a visibility measures a single sky mode, with a delta-function footprint in the Fourier plane. For a real instrument with a finite primary beam and bandpass, modes are mixed into an individual visibility measurement due to leakage. For a small FOV, the corresponding beam in Fourier space is broad, because the beam convolution distributes the power from adjacent modes. This beam footprint determines the coherence length of Fourier modes, as demonstrated schematically in Figure \ref{beam_footprint}.
Here, the beam footprint from individual visibility measurements determines the joint sky information available on a given angular scale.

The thermal noise is correlated only for \textit{visibilities} (i.e., the measurement basis), while the sky power is correlated in the sky basis. The effective number of visibilities contributing to a given mode, $N_{{\rm eff}, i} = 1/[\left( \boldsymbol{B}^{\dagger}\boldsymbol{B} \right)^{-1}]_{ii}$, provides the coherence of the data with the sky. The same is not true for the cosmic variance, where the power is the same for identical sky \textit{modes}, and uncorrelated otherwise (under the assumption of a Gaussian random field). The diagonal components of $\left(\boldsymbol{B}^\dagger \boldsymbol{B}\right)$ integrate visibilities that contribute coherently, while the off-diagonal components encode the correlations between sky modes introduced by the finite Fourier beam.


\subsection{Incoherent combination of information: power covariance}
To reduce cosmic variance, and further reduce the thermal noise, we incoherently combine information from modes within some annulus of constant $k = \sqrt{u^2+v^2}$. The beam footprint, $(\boldsymbol{B}^\dagger\boldsymbol{B})$, encodes the correlation between modes within this annulus, as shown schematically in Figure \ref{beam_footprint} by the blue annular region. 
We form the power \textit{sample} in mode $\overline{u_i}$ by squaring the sky signal and normalising by the instantaneous observation volume,
\begin{equation}
P(\overline{u_i}) = \frac{S^{\dagger}(\overline{u_i}) S(\overline{u_j})}{ V_u},
\end{equation}
and then form the maximum likelihood (ML) incoherent power average over $uv$-modes contributing to an annulus at ($\overline{k}$, $\overline{k}+\Delta{\overline{k}}$). For \textit{independent} modes, this reduces to \citep{mcquinn06}:
\begin{equation}
P(\overline{k}) = \frac{ \displaystyle\sum_{\overline{u_i} \in k} N(\overline{u_i}) P(\overline{u_i})}{ \displaystyle\sum_{\overline{u_i} \in k} N(\overline{u_i}) },
\label{incoherent}
\end{equation}
where $N(\overline{u_i})$ is the number of measurements contributing to mode $\overline{u_i}$.

For a general covariance, we can derive the correct ML estimate. The covariance matrix of the \textit{power} is found by considering the covariance matrix of the sky signal, $\boldsymbol{C}_s$, and the propagation of errors for a quadratic form\footnote[1]{For a 2D power spectrum, the deconvolved sky signal is Fourier transformed along the frequency dimension (line-of-sight):
\begin{equation}
\overline{S}(\overline{k}) = F (\boldsymbol{B}^\dagger\boldsymbol{B})^{-1}\boldsymbol{B}^\dagger \overline{V},
\end{equation}
where $F$ denotes the line-of-sight Fourier transform. The power covariance matrix is then:
\begin{equation}
\boldsymbol{C}_p V_k^2 = \left(\sigma^2 \left( \boldsymbol{F}^\dagger \boldsymbol{B}^{\dagger}\boldsymbol{B} \boldsymbol{F} \right)^{-1}_{ij} + \boldsymbol{P}^\prime \right)^\dagger \left(\sigma^2 \left( \boldsymbol{F}^\dagger \boldsymbol{B}^{\dagger}\boldsymbol{B} \boldsymbol{F} \right)^{-1}_{ij} + \boldsymbol{P}^\prime\right),
\end{equation}
where $V_k$ is the observation volume (sr.Hz). For independent frequency channels, the Fourier transform sums the noise contributions from each channel, and we form the approximation;
\begin{equation}
\left( \boldsymbol{F}^\dagger \boldsymbol{B}^{\dagger}\boldsymbol{B} \boldsymbol{F} \right)^{-1} \simeq \left( \displaystyle\sum_{i=1}^{N_{\rm ch}} \boldsymbol{B}^{\dagger}\boldsymbol{B} \right)^{-1}.
\end{equation}}
:
\begin{eqnarray}
\boldsymbol{C}_p &=& \frac{\boldsymbol{C}_s^\dagger \boldsymbol{C}_s}{ V_u^2}\\
&=& \frac{1}{V_u^2}\left(\sigma^2 \left( \boldsymbol{B}^{\dagger}\boldsymbol{B} \right)^{-1}_{ij} + \boldsymbol{P}\Omega_u \right)^2.
\end{eqnarray}
The ML solution for the power in mode $k$ is found by considering the likelihood function for the power data. For simplicity we assume that the thermal noise dominates the sky power in each coherent mode (a reasonable assumption for the current generation of experiments, which are predicted to detect the EoR signal only after incoherent averaging), and approximate the ML estimate for the power in mode $k$ as:
\begin{equation}
\hat{\boldsymbol{P}}_k = \frac{\displaystyle\sum_{i\in k}  \boldsymbol{C}_p^{-1} \overline{P}}{\displaystyle\sum_{i\in k}  \boldsymbol{C}_p^{-1} }.
\label{power_ml}
\end{equation}
Finally, to obtain the uncertainty in the $k$-mode power estimate, $\hat{P}_k$, we compute the variance of Equation \ref{power_ml}:
\begin{equation}
\Delta{P}_k^2 = \frac{1}{\displaystyle\sum_{i\in k} \boldsymbol{C}_p^{-1}} = \frac{1}{V_u^2 \displaystyle\sum_{i\in k} \boldsymbol{C}_s^{-1}\boldsymbol{C}_s^{-1}},
\label{uncert_ml}
\end{equation}
where the sum combines all of the elements of the inverted covariance matrix that contribute to that mode, $u_i\in{k}$. 
To invert the power covariance matrix, we note that it is composed of a matrix inverse [$(B^\dagger{B})^{-1}$] and diagonal signal matrix ($P^\prime$). In order to avoid inverting the beam weights matrix, we use a Taylor expansion:
\begin{equation}
(A + B)^{-1} \approx A^{-1} - A^{-1}BA^{-1} + A^{-1}BA^{-1}BA^{-1} - \cdots,
\end{equation}
giving;
\begin{equation}
\boldsymbol{C}_p^{-1} \approx V_u^2 \left( \displaystyle\sum_{i=1}^{4} \frac{1}{\sigma^{2i}} (-1)^{i+1} \Omega_u^{2(i-1)} \boldsymbol{P}_k^{i-1} \left( \boldsymbol{B}^{\dagger}\boldsymbol{B} \right)^i \right)^2,
\label{inverse_power_cov}
\end{equation}
where we have approximated the coherent power in each mode $P(u_i) \approx P_k$ for $i \in k$. Note that this expression makes the computation of the final power variance straightforward, because it \textit{does not require inversion of a large matrix}.


Together, equations \ref{uncert_ml} and \ref{inverse_power_cov} provide the quantity in which we are interested; the uncertainty in the measured power in mode $k$. These expressions describe the normal ML incoherent power (equation \ref{incoherent}). For addition of independent modes, this expression reduces to the approximate form from \citet{mcquinn06} (their equation 24).

\section{Observing modes}
\label{beam_section}
We are interested in understanding the uncertainty in computing the power at a given scale, for different observing schemes. We consider three observing schemes:
\begin{enumerate}
\item Tracked --- beam pointing centre matches the phase centre, which have a constant sky position;
\item Zenith drift --- pointing centre = zenith, phase centre = EoR field centre. (Note that this observing scheme may have the phase centre outside of the main lobe of the primary beam. For the MWA, this occurs in $\sim$2 hours);
\item Drift$+$shift --- fixed pointing centres, in steps across the sky, phase centre = EoR field centre (approximates a tracked scan).
\end{enumerate}
The key differences here lie in the influence of the antenna primary beam pattern on the sampling of the sky, and the effect this has on the coherence of measured data. 

The beam is the Fourier transform of the primary beam (PB) pattern on the sky, relative to the phase centre of the observation (for antennas with different beams, in Fourier-space it is the convolution of the antenna illumination pattern for each --- for simplicity here, we will consider identical beams).
\begin{enumerate}
\item {\bf Tracked}: a tracked observation is characterised by the same pointing and phase centres through the observation. The Fourier beam is simply the Fourier transform of the real-space PB, and is real-valued for a symmetric PB. In general, the Fourier beam can evolve through the observation, according to the PB pattern as a function of pointing centre, and this can be incorporated into the beam patterns used for those pointings. The tracked Fourier beam, $B_{tr}(\overline{u})$ is given by;
\begin{eqnarray}
B_{tr}(\overline{u}) &=& \mathcal{FT}(B(\overline{l}))\\\nonumber
&=& \int B(\overline{l}) \exp{(-2\pi{i} \overline{u} \cdot \overline{l})}.
\end{eqnarray}
For this mode, the observation volume is given by:
\begin{equation}
\Omega_u \equiv V_u = \displaystyle\int d^3\overline{u} |B(\overline{u})|^2.
\end{equation}
\item {\bf Zenith drift}: the zenith drift observation is characterised by a constant pointing centre (zenith) and stable PB, and constant sky phase centre, which changes with respect to the pointing. The PB pattern therefore rotates relative to the sky. A rotation in real-space is represented by a phase shift in Fourier space, leading to the zenith drift Fourier beam, $B_z(\overline{u};\theta)$,
\begin{eqnarray}
B_{z}(\overline{u};\theta) &=& \mathcal{FT}(B(\overline{l}-\Delta{\theta}))\\\nonumber
&=& \int B(\overline{l}-\Delta{\theta}) \exp{(-2\pi{i} \overline{u} \cdot \overline{l})}\\\nonumber
&=& \exp{(-2\pi{i} \overline{u} \cdot \Delta{\theta})} B_{tr}(\overline{u})\\\nonumber
&\equiv& \overline{H}({\theta}) B_{tr}(\overline{u}),
\end{eqnarray}
where $\Delta{\theta}=\Delta\overline{\theta}(t)$ is the time-dependent vector offset between pointing and phase centres, and we have defined $\overline{H}(\theta)$, a vector encoding the phase shifts at each angular scale. When considering the evolution of the angular offset with time, this becomes a matrix, $\boldsymbol{H}(\theta)$. This produces a scale-dependent ($u$-dependent) phase rotation in the beam, whereby the largest angular scales on the sky are least affected. For a symmetric PB, this introduces imaginary components into the beam. Over the course of an observation, the telescope beam evolves, such that the integrated beam footprint is;
\begin{equation}
\boldsymbol{B}^\dagger\boldsymbol{B} = \int \boldsymbol{H}^\dagger(\theta) (\boldsymbol{B}^\dagger\boldsymbol{B})(\theta) \boldsymbol{H}(\theta) d\theta.
\end{equation}
The observation volume is then,
\begin{eqnarray}
\Omega_u &=& \displaystyle\int d\overline{u}^\prime B_1(\overline{u}-\overline{u}^\prime) B_2^\ast(\overline{v}-\overline{u}^\prime)  \\\nonumber
&=& \displaystyle\int d^3\overline{u} \left( \int d\theta \boldsymbol{H}^\dagger(\theta) (\boldsymbol{B}^\dagger\boldsymbol{B})(\theta) \boldsymbol{H}(\theta) \right)^2 \\\nonumber
&\simeq& \displaystyle\int d^3\overline{u} \left|B(\overline{u})\right|^2 {\rm sinc}(2\pi\overline{\theta}_{\rm max} \cdot \overline{u}) ^2,
\end{eqnarray}
where the approximation comes from considering the beam footprint to be independent of angular offset, $\theta$. For the purposes of this calculation (to compute $\Omega_u$), this approximation is appropriate.
\item {\bf Drift$+$shift}: the drift$+$shift observation is an intermediate protocol between the tracked and pure-drift scans. It is characterised by a constant sky phase centre throughout the observation, but periods of drift at fixed pointing centres. For the MWA, the field drifts for 30 minutes ($\sim$8 degrees) before re-pointing.
\end{enumerate}


\section{Simulations}
\label{mwa_experiment}
In accordance with the EoR experiment description in \citet{bowman13_mwascience}, a 900-hour observation is simulated for the MWA (6 hours per night), with the phase centre set to a fixed sky position at a declination of  $\delta{=-}$30 degrees. We use a position- and frequency-dependent analytic approximation to the MWA primary beam, including the effects of the ground screen and phasing of 16 individual dipoles. The $uv$ coverage of visibilities is computed for hypothetical one minute integrations and 200~kHz spectral channels over a total bandwidth of 8~MHz.  A sky Fourier grid is created with a sampling of $\Delta{u}=\Delta{v}=0.5$, to Nyquist sample a region of sky equivalent to 6 hours of drift. To reduce computational load, a maximum $uv$ distance of $|\overline{u}|_{\rm max}=80$ ($l_{\rm max}=500$, $k_{\rm max}=0.077$ $h$Mpc$^{-1}$) is chosen. Baselines outside of this range are ignored, yielding a significantly reduced dataset. The drift$+$shift has 12 pointing centres over 6~hours, with 30 minutes at each pointing. The thermal noise per visibility, $\sigma$, is computed for a sky-dominated system temperature, T$_{\rm sys}$ = 440~K, spectral resolution of $\Delta\nu=$~200~kHz, and two polarizations, yielding:
\begin{equation}
\sigma = \frac{T_{\rm sys}\lambda^2\Delta\nu}{A_e\sqrt{2\Delta\nu\Delta{t}}} \simeq 5,000 \,\,{\rm K.sr.Hz}.
\end{equation}
This value is used for each visibility measured by the experiment, with an integrated $\sim{10}^{10}$ visibilities combined over the course of the experiment (10 second cadence, with 128 tiles and 900 hours total observing time).

\begin{figure}
\begin{center}
\includegraphics[scale=0.5]{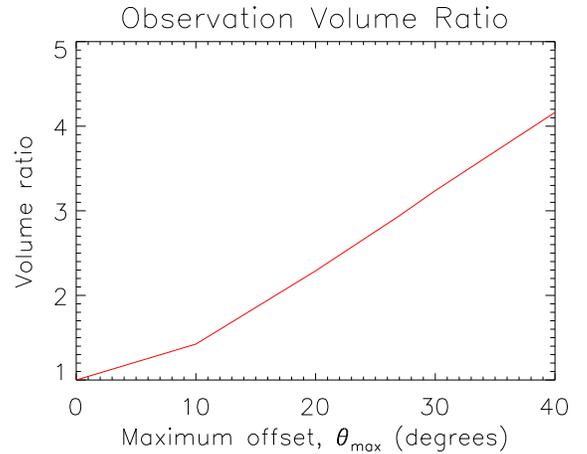}
\caption{Observation volume, $\Omega_u$, for a zenith drift scan, relative to a tracked scan for the MWA, as a function of the maximum offset between the pointing centre and phase centre.}\label{obs_volume}
\end{center}
\end{figure}

Equation \ref{c_s} has two terms to compute. For the thermal noise term, we compute the beam vector, $\overline{B}_V$, for each single visibility, based on its $uv$ location and the observing mode, and compute the outer product to increment the beam matrix. \textit{Viz}:
\begin{equation}
(\boldsymbol{B}^\dagger\boldsymbol{B})_{i+1} = (\boldsymbol{B}^\dagger\boldsymbol{B})_{i} + (\overline{B}_V^\dagger{\overline{B}_V}).
\end{equation}
For the cosmic variance, the observation volume, $\Omega_u$, in equation \ref{c_s} is computed by performing the integral over the integrated beam footprint. Figure \ref{obs_volume} shows the observation volume, relative to a tracked scan, for the zenith drift and drift$+$shift modes, as a function of maximum angle between phase and pointing centres.
For this experiment, we consider a maximum angle of 27 (4) degrees for drift (drift$+$shift), with a corresponding volume ratio of $\Omega_u/\Omega_u({\rm track})=2.9$ (1.3).

We compute the uncertainty in the 2D power, $P_k$ (mK$^2$.Mpc$^3{h}^{-3}$) by application of equations \ref{uncert_ml} and \ref{inverse_power_cov}, for linearly-spaced angular $k$-bins with a width of $\Delta{k_\bot}=1.5\times{10}^{-3}h{\rm Mpc}^{-1}$. We then compute the 1D power spectrum from the same simulations by forming the power for each $k$;
\begin{equation}
k = \sqrt{k_\bot^2 + k_\parallel^2}.
\end{equation}

\subsection{Foregrounds}\label{foregrounds}
We explicitly included the contribution from unsubtracted foregrounds in the signal covariance matrix, but have, as yet, not distinguished between cosmological signal and foregrounds in the noise analysis. Most current and future experiments aim to remove the strongest of these sources to high precision (others, such as PAPER, \citet{parsons10}, try to work in a region of parameter space where foreground contamination is minimal), but emission will remain due to unremoved sources (often below the confusion limit) and imprecise source subtraction \citep[e.g., ][]{trott12}. The information available in the data limits the level and precision to which foregrounds can be removed. Although foregrounds are contaminants to the signal of interest, they have a different signature in the 2D power spectrum parameter space. Previous theoretical work has demonstrated the expected existence of a wedge feature, which is the signature of imprecisely- and un-subtracted smooth-spectrum point sources \citep{bowman09,datta10,trott12,vedantham12,morales12,thyagarajan13}. For an achromatic instrument, spectrally-smooth foregrounds occupy the lowest $k_\parallel$ modes (flat-spectrum sources strictly occupy only the $k_\parallel=0$ mode). However, interferometers are inherently chromatic, with each physical baseline sampling signal at different wavenumbers for each spectral channel. This yields the wedge-like structure in the 2D power spectrum, where larger $k_\perp$ modes have greater leakage into higher $k_\parallel$ modes. For the small $k_\perp$ modes studied in this work, and the coarse spectral sampling, the wedge occupies only the $k_\parallel=0$ mode. To avoid the contribution of foregrounds, in the foregoing analysis, we omit signal in this mode.

\section{Results}
\begin{figure}
\begin{center}
\includegraphics[scale=0.5]{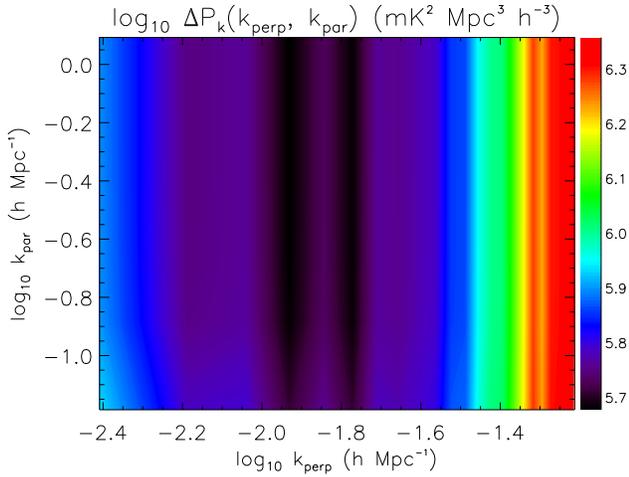}
\caption{Uncertainty in power for a zenith drift observation (mK$^2$~Mpc$^3$~$h^{-3}$ ).}\label{cl}
\end{center}
\end{figure}

\begin{figure}
\begin{center}
\includegraphics[scale=0.5]{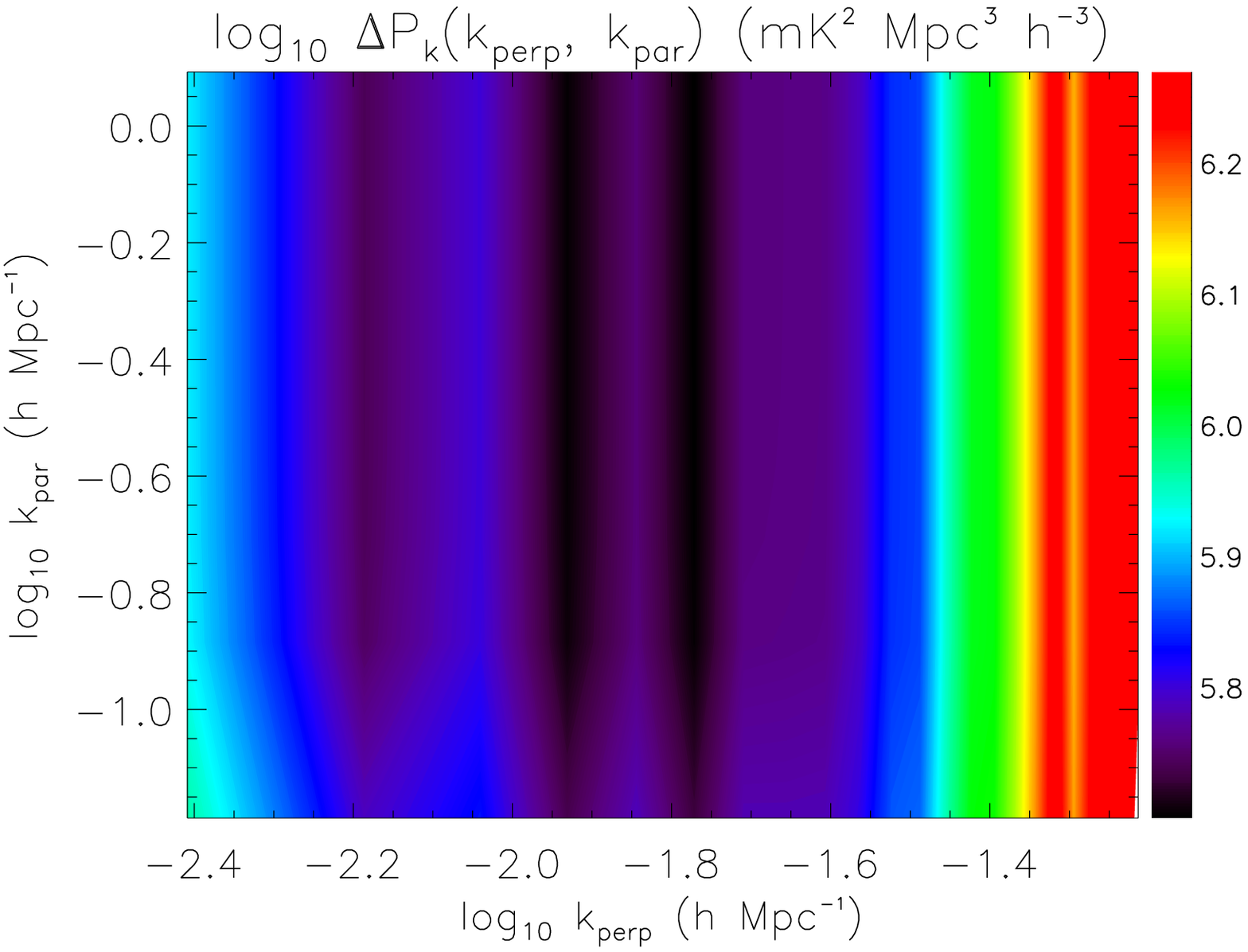}
\caption{Uncertainty in power for a drift$+$shift observation (mK$^2$~Mpc$^3$~$h^{-3}$ ).}\label{cl_dim}
\end{center}
\end{figure}

\begin{figure}
\begin{center}
\includegraphics[scale=0.5]{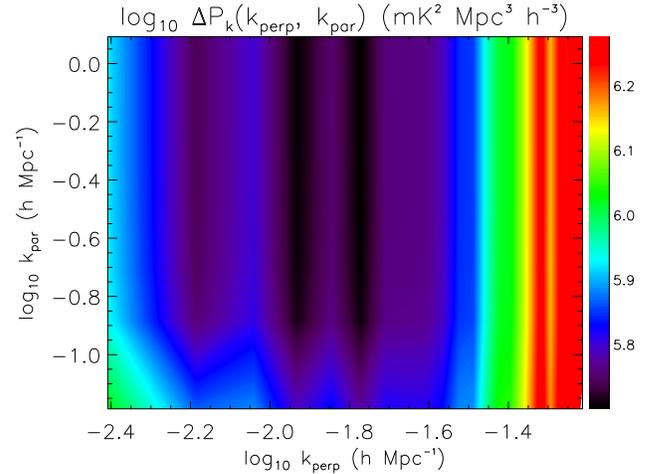}
\caption{Uncertainty in power for a tracked observation (mK$^2$~Mpc$^3$~$h^{-3}$ ).}\label{cl_dim_track}
\end{center}
\end{figure}

\begin{figure}
\begin{center}
\includegraphics[scale=0.5]{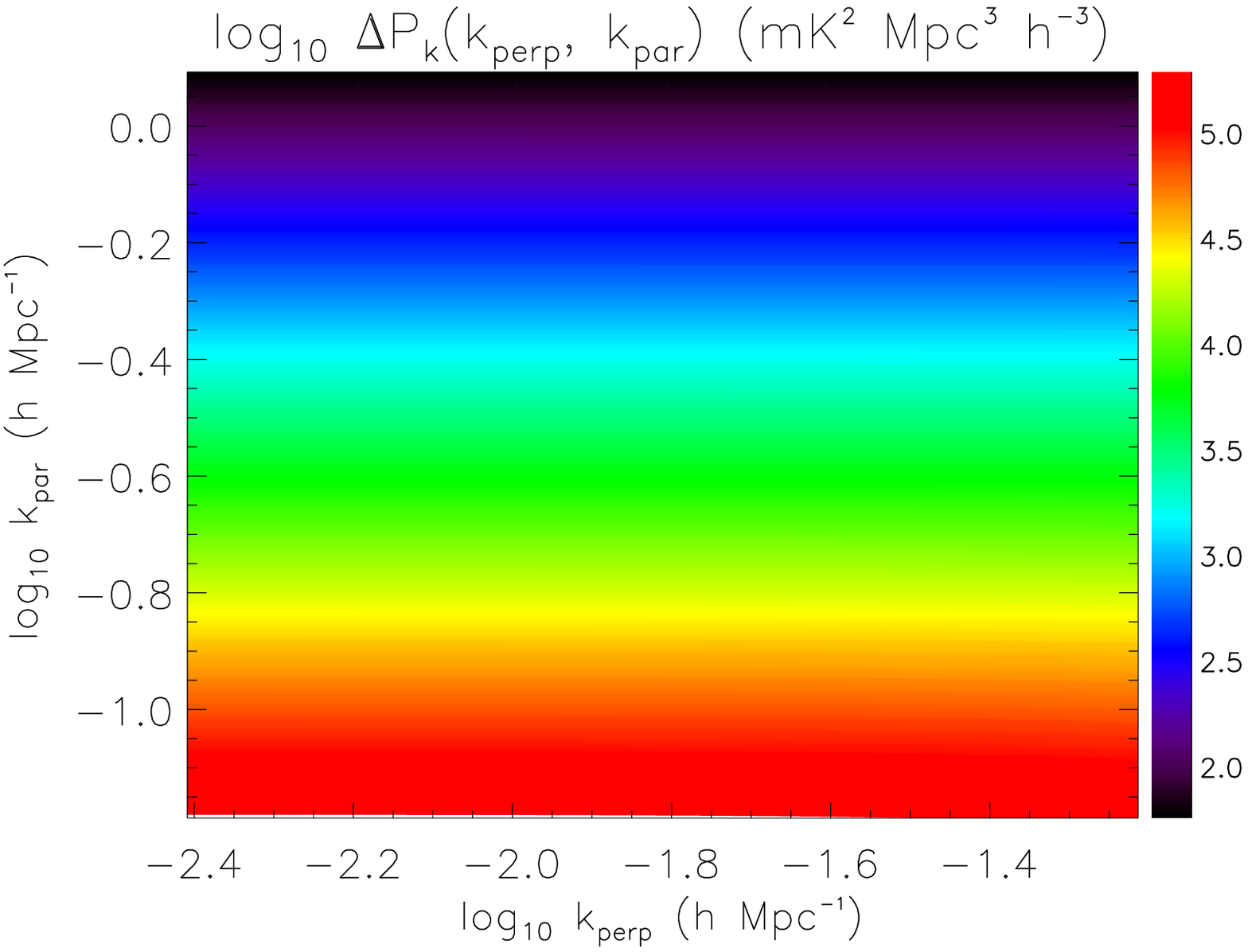}
\caption{Reference input 21~cm emission signal, assuming a neutral IGM at $z$=8 \citep{eisenstein99,furlanetto06}.}\label{cl_dim_zoom}
\end{center}
\end{figure}

\begin{figure}
\begin{center}
\includegraphics[scale=0.5]{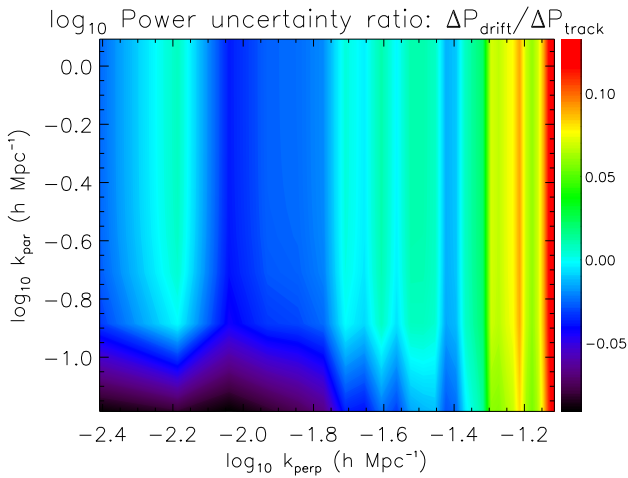}
\caption{Ratio of noise uncertainty in each mode, log$_{10}$~$\Delta{P_{\rm drift}}/\Delta{P_{\rm dns}}$.}\label{ratio}
\end{center}
\end{figure}

Figures \ref{cl}, \ref{cl_dim} and \ref{cl_dim_track} display the uncertainty in power for a zenith drift, drift$+$shift, and tracked mode, respectively, in the 2D power spectrum.
For reference, Figure \ref{cl_dim_zoom} displays the input 21~cm power spectrum used in the simulations \citep[$z$=8, neutral IGM, including line-of-sight anisotropy;][]{eisenstein99,furlanetto06}.
The overall distribution of noise traces the distribution of baseline lengths for the MWA, with fewer long baselines relative to short (log$_{10}k_\bot<-1.8$). The results also reflect our expectations at these large angular scales, where thermal noise is high (few very short baselines). At low $k_\parallel$, the expected signal strength is highest, and cosmic variance contributes noticeable noise uncertainty for the drift$+$shift and tracked modes. This additional noise is reduced for the zenith drift scan, where the cosmic variance is decreased by the additional sky coverage. Figure \ref{ratio} displays the log$_{10}$ ratio of noise uncertainty for the zenith drift mode relative to the tracked mode.
The effect of the reduced cosmic variance for the zenith drift scan mode is clearly evident at low $k_\parallel$, but the overall improvement in these modes is less than a factor of two in noise uncertainty. At large $k_\bot$, the drift$+$shift and tracked modes perform better because the experiment is thermal noise-dominated and there is less coherence for the zenith drift scan mode.

\begin{figure}
\begin{center}
\includegraphics[scale=0.5]{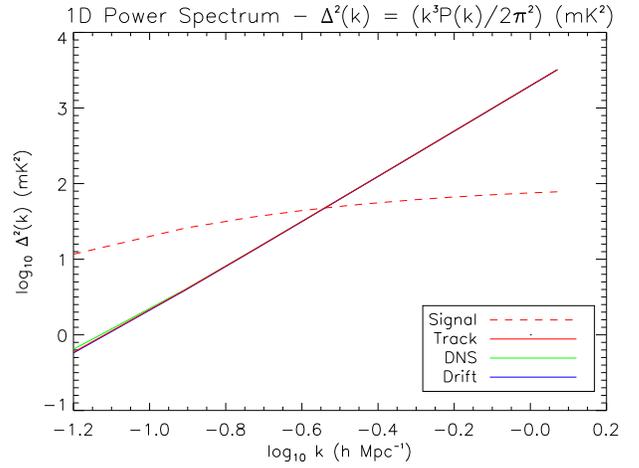}
\caption{Expected dimensionless power spectrum (signal), and uncertainty for the track, drift$+$shift and drift observing modes.}\label{spherical}
\end{center}
\end{figure}

Figure \ref{spherical} shows the uncertainty in the 1D dimensionless power spectrum, where,
\begin{equation}
\Delta^2(k) = \frac{k^3P(k)}{2\pi^2}.
\end{equation}
The expected 21~cm signal is shown for comparison.
The S/N exceeds unity at large scales. The three observing modes show visually similar results, except at large scales where the zenith drift scan exhibits marginally improved performance. This is expected given the results of the 2D power spectrum; because only small $k_\bot$ modes are considered here, $k$ is dominated by $k_\parallel$ for larger modes. This is where there is reduced cosmic variance, and the two observing modes show comparable results (compare with Figures \ref{cl}--\ref{cl_dim_track}).

The dimensionless power spectrum can be compared with the expectations for MWA sensitivity to EoR statistical estimation computed in \citet{beardsley13}. They compute the S/N on the amplitude [$\Delta^2(k_p)$] and slope ($\alpha$) of the dimensionless power spectrum for a single field and 450 hours of observing where \citep{lidz08},
\begin{equation}
{\rm ln}\Delta^2(k) = {\rm ln}\Delta^2(k_p) + \alpha{\rm ln}\frac{k}{k_p},
\end{equation}
and $k_p = 0.06$~$h$Mpc$^{-1}$ is the pivot wavenumber. We compute the Cramer-Rao Bound \citep[CRB, ][]{kay98} on the minimum uncertainty on estimation of these parameters for an efficient, unbiased estimator, and these data. The S/N on the parameters for 900 hours of observing are shown in Table \ref{snr_table}. 
\begin{table}
\begin{center}
\begin{tabular}{|c||c|c|c||c|}
\hline 
 & Drift & Drift$+$shift & Track & SNR$_D$/SNR$_T$ \\ 
\hline 
S/N$_{\Delta^2_p}$ & 9.60 & 9.51 & 9.31 & 1.03 \\ 
\hline 
S/N$_{\alpha}$ &  6.60 & 6.54 & 6.51 & 1.01 \\
\hline
\end{tabular} 
\caption{Signal-to-noise ratios on estimation of the amplitude and slope of the spherically-averaged power spectrum, for the three observing strategies.} \label{snr_table}
\end{center}
\end{table}
These can be compared with the quantities from \citet{beardsley13} for one field and the same observing time, S/N$_{\Delta^2_p} =10.5$, S/N$_{\alpha} =8.2$. Thus, we predict slightly reduced precision on the value of the slope and the amplitude. There are differences in the methodology, input 21~cm signal, and binning of spectra used in the two analyses, but the primary difference lies in the correct accounting for the coherence of measurements in this work. Importantly, both this work and \citet{beardsley13} predict detectability of the signal with 900 hours of observing.

More interesting is the direct comparison of the three observing modes in this work, because these are computed using exactly the same experimental design, input signal and power spectrum binning. We predict improvement in detectability of 1\% (3\%) for the slope (amplitude) using the zenith drift mode over the tracked mode, for a drift scan encompassing $\pm$27 degrees of zenith. This translates into direct reduction of required observing time for the same level of detectability. I.e., a zenith drift scan with 875 hours of observing can achieve similar precision on the slope, and better on the amplitude, to a tracked scan of 900 hours duration.

\section{Discussion and conclusions}
\label{discussion}

Observing strategy is a key consideration for current and future EoR estimation experiments with aperture arrays. The complexity of the antennas motivates us to consider system stability and instrument calibratability when designing our science experiments. In this work we considered three qualitatively different approaches to measuring the EoR signal with the redshifted 21~cm emission line, and compared them on the basis of noise uncertainty in the 2D power spectrum. Having developed a general framework, we applied it to the EoR experiment underway with the MWA telescope. The analysis suggests that, for this experiment, there is up to a factor of two difference in power uncertainty across the 2D parameter space of spatial modes. Interestingly, the zenith drift scan shows improved performance in the regions of interest for the EoR signal, due to its effective reduction of cosmic variance and higher sensitivity. For the dimensionless 1D power spectrum, we predict marginal gains for using the zenith drift scan, which can be directly translated into a marginal reduction in observing time required to reach the same level of parameter estimation precision. More critically, the comparable performance of the zenith drift scan makes it an attractive option for instruments where calibration stability is a primary concern.

Although the gains in estimation SNR derived here are modest, they refer specifically to the case in question: the 900-hour EoR experiment being undertaken with the MWA. An increase in time-on-sky for the same experiment would favour the drift mode, because it would further reduce thermal noise and increase the relative importance of the cosmic variance. Conversely, less time-on-sky would still lead to a significant estimate of the 1D parameters of slope and amplitude, but thermal noise would be more important and the tracked scan may be favoured. Indeed, the choice of 900 hours is partly due to that corresponding to an approximate balance of these noise terms.

It is worth noting that there may be other differences between these observing modes than simply the coherence of measurements. For example, the feature that makes the drift scan attractive (its large observing volume) may also be a detriment, where it is more difficult to find regions of the sky without strong signal contamination (e.g., the Galactic centre) in a sidelobe of the primary beam. The impact of the precise choice of experimental location depends greatly on the location, and this cannot be assessed in a general way, and has not been explored here. Ultimately, any EoR experiment requires a careful tuning of the experimental conditions to yield the maximum scientific gain. In this work, we have demonstrated that the zenith drift scan is marginally advantageous for signal estimation from the noise perspective alone.

For future experiments with higher sensitivity \citep[e.g., LEDA, SKA-Low;][]{greenhill12_leda,schilizzi10}, coherent reduction of thermal noise will be less important, and the drift scan mode may show additional benefits for reducing cosmic variance. Coupled with the benefits of improved system stability and calibratability of this mode, zenith drift scans may become the mode of choice for statistical estimation of the EoR signal.

\begin{acknowledgements}
The Centre for All-sky Astrophysics is an Australian Research Council Centre of Excellence, funded by grant CE110001020. The International Centre for Radio Astronomy Research (ICRAR) is a Joint Venture between Curtin University and the University of Western Australia, funded by the State Government of Western Australia and the Joint Venture partners. CMT would like to thank Adam Beardsley for providing details of his MWA sensitivity calculations, and also the anonymous referee, Ron Ekers, Randall Wayth, Jonnie Pober, Nithya Thyagarajan, and the MWA EoR collaboration, for fruitful discussions and substantial improvement of the manuscript.
\end{acknowledgements}

\bibliographystyle{apj}
\bibliography{pubs}

\end{document}